# Governance-Constrained Agentic AI: Blockchain-Enforced Human Oversight for Safety-Critical Wildfire Monitoring

Ali Akarma [1,*], Toqeer Ali Syed [1], Salman Jan [2], Hammad Muneer [3] and Abdul Khadar Jilani [4]

1. AI Center, Faculty of Computer and Information Systems, Islamic University of Madinah, Madinah 42351, Saudi Arabia; 443059463@stu.iu.edu.sa (A.A.); toqeer@iu.edu.sa (T.A.S.)
2. Faculty of Computer Studies, Arab Open University-Bahrain, Manama 32038, Bahrain. salman.jan@aou.org.bh (S.J.)
3. Department of Computer Science, The Islamia University of Bahawalpur, Bahawalpur 63100, Pakistan. S21bdocs1m01009@iub.edu.pk (H.M.)
4. College of Computer Studies, University of Technology Bahrain, Sitra 18041, Bahrain; a.jilani@utb.edu.bh (A.K.J.)

* Correspondence: 443059463@stu.iu.edu.sa.

**Abstract**

The AI-based sensing and autonomous monitoring have become the main components of wildfire early detection, but current systems do not provide adaptive inter-agent coordination, structurally defined human control, and cryptographically verifiable responsibility. Purely autonomous alert dissemination in the context of safety critical disasters poses threats of false alarming, governance failure and lack of trust in the system. This paper provides a blockchain-based governance-conscious agentic AI architecture of trusted wildfire early warning. The monitoring of wildfires is modeled as a constrained partially observable Markov decision process (POMDP) that accounts for the detection latency, false alarms reduction and resource consumption with clear governance constraints. Hierarchical multi-agent coordination means dynamic risk-adaptive reallocation of unmanned aerial vehicles (UAVs). With risk-adaptive policies, a permissioned blockchain layer sets mandatory human-authorization as a state-transition invariant as a smart contract. We build formal assurances such as integrity of alerts, human control, non-repudiation and limited detection latency assumptions of Byzantine fault. Security analysis shows that it is resistant to alert injections, replays, and tampering attacks. High-fidelity simulation environment experimental evaluation of governance enforcement demonstrates that it presents limited operational overhead and decreases false public alerts and maintains adaptive detection performance. This work is a step towards a principled design paradigm of reliable AI systems by incorporating accountability into the agentic control loop of disaster intelligence systems that demand safety in their application.

## 1. Introduction

Wildfires are safety-critical, high uncertainty incidents which should be detected and adaptively monitored within seconds and communicated to the population through reli-able processes. The rise in climate variability and the spread of human-wild land interfaces have also elevated the intensity and frequency of wildfire events, which is why the early warning systems will play a crucial role in reducing the losses to ecology and human lives as a whole [1]. The recent development of deep learning, satellite imaging, and unmanned aerial vehicles (UAVs) has enhanced automated anomaly detection, but most current systems are reactive with a fixed patrol plan and without any enforceable governance system in place, so far [2]. Purely autonomous alert dissemination creates risks of false alert, untimely broadcast, and loss of trust in people in safety-critical disaster conditions, which are unique to autonomous disaster warning systems [3]. In this context, "Agentic" is used to mean systems, which are goal oriented, closed loop perception-planning-action, adaptive policy revision in the face of uncertainty and explicit reasoning about future action consequences.

The difficulty is a more general agentic artificial intelligence problem; how can one create adaptive autonomous systems that are both efficient and, crucially, have human overseers and accountability? The existing pipelines of wildfire monitoring place significant focus on the accuracy of perception but seldom combine (i) a coordinated multi-agent resource distribution in the context of uncertainty, (ii) cryptographically verifiable audit trail, and (iii) human validation through the structural means before the issue is reported to the public.

In this paper, the problem of wildfire monitoring is defined as a constrained sequential decision-making problem of governance with partial observability. The system needs to learn from the heterogeneous sensing inputs the latent ignition states and dynamically assign UAV resources to reduce the detection latency and false public alerts. Simultaneously, the alerts dissemination should be conditional on the enforceable human authorization. We use a governance-limited agentic architecture, combining adaptive multi-agent coordination with blockchain-enforced human-in-the-loop (HITL) validation to meet these requirements [4,5].

The key contributions of this work are as follows:
- A decision-theoretic model of adaptive monitoring of wildfires constrained by resource and governance limitations.
- A hierarchical multi-agent system of coordination that facilitates risk-sensitive UAV redeployment.
- An administrative layer based on a blockchain, which cryptographically ensures that humans must approve mandatorily before alerting is sent.
- An experimental assessment of less detection latency and false public alerts with limited governance overhead.

## 2. Related Work

The paper cuts across four domains AI-based detection of wildfires, multi-agent coordination in the presence of uncertainty, human-in-the-loop (HITL) safety-critical AI, and blockchain-based accountability. Although both have grown separately, they have not been integrated fully under governance restrictions that are enforced formally.

Recent systems use convolutional neural networks and transformer architectures and multimodal fusion on satellite, UAV thermal feed, and ground sensors [2]. Though the schemes enhance accuracy of anomaly detection, they are normally perception-driven pipelines that generate alerts on a threshold basis [6]. There is a general lack of adaptive resource allocation and cryptographically enforced human validation to restrict formal governance assurances.

The multi-agent systems (MAS) study has come up with auction-based allocation, distributed reinforcement learning, decentralized POMDP solvers, and coverage control of adaptive patrol under uncertainty [7]. These approaches maximize coverage efficiency or latency but often do not include any obligatory external authorization limitation in the decision-making-process, a critical condition in safety-critical settings [8]. HITL architecture minimizes automation errors in high-risk areas, but their control is often carried out in a procedural manner at the interface level [9]. Human authorization is rarely encoded in any system as a state-transition invariant which permits autonomous execution only when the approval has been verified.

Permitted blockchains are tampered loggers with Byzantine fault tolerance [5]. They are primarily applied as data provenance or post hoc auditing [10] in disaster monitoring and IoT environments to control system outputs, as opposed to enforcement controls. Contrarily, this work integrates adaptive coordination, a decision-theoretic model that is constrained, cryptographically governed and mandatory human authorization into a single architecture. The framework provides adaptive autonomy through authorization restrictions to a constrained POMDP and alert release through smart contracts, whose integrity and oversight is proven.

## 3. System Model and Threat Assumptions

Our suggested wildfire monitoring structure will be a decentralized cyber-physical architecture, consisting of sensing, autonomous coordination, blockchain governance, and human validation.

*3.1. System Architecture*

The system consists of the following components:

**Physical IoT Sensing Layer:** A collection of heterogeneous sensors $\mathcal{S}$ composed of UAV thermal cameras, ground IoT sensors, satellite feeds, and meteorological sources of data produce timestamped observations of the environmental conditions. The outputs of the sensor are sent towards a coordination engine to become probabilistically fused out [11].

**Digital Twin Layer:** A dynamic wildfire risk digital twin keeps a 3D representation of the environment of the area under observation continuously updated. This model incorporates multi-modal sensing data to conduct predictive simulation, scenario analysis and risk forecasting. The digital twin assists the agentic coordination engine with the spatial risk priors and propagation estimates, which are utilized in the patrol reallocation and anomaly verification.

**Agentic UAV Layer:** Adaptive patrol and anomaly verification is done by a fleet of autonomous UAV agents $\mathcal{U}$. The state variables of every UAV are position and battery level and the redeployment commands are executed under a centralized decision module identification system based on a centralized decision module [2].

**Coordination Engine:** The agentic AI layer comprises of specialized agents such as (i) a fire risk prediction agent to dynamically estimate the risk (ii) an UAV coordination agent to allocate adaptive patrols and (iii) a verification and confidence agent to perform a multi-stage anomaly validation. These modules run on a reasoning engine that has an execution of perception, contextual reasoning, choosing an action and online learning. The system has a belief state of possible ignition areas by the use of multi-modal sensor fusion. According to this belief, the system chooses such actions as UAV redeployment, multi-stage verification, or the generation of alerts requests.

**Blockchain Governance Layer:** Detected anomalies are serialized as transactions containing event identifiers, timestamps, geospatial references, confidence scores, and hashes of supporting evidence [5]. These are transactions which are sent to a permissioned blockchain network $\mathcal{V}$ which is run by authorized agencies. Smart contracts ensure that the public dissemination of alerts is only allowed to pass when specifics of governance have been met as a predetermined condition of dissemination [12].

**Human Validation Authority:** Human operators are allowed to communicate via a dashboard interface that displays explainable artificial intelligence output, anomaly confidence, and evidence. An approval and safety validation gate is used to mediate human override or authorization decisions before the distribution of alert. This only allows public diffusion after cryptographic human approval, which is implemented at the smart-contract level of cryptography [13].

The overall architecture is illustrated in Fig. 1.

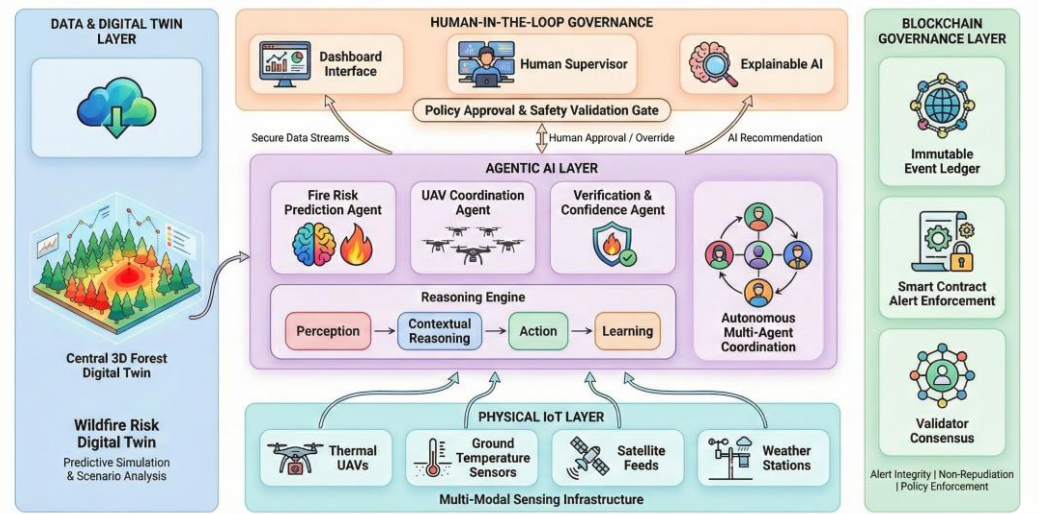

**Figure 1.** Architecture of an agentic AI–driven wildfire risk management system integrating digital twins, multi-agent coordination, human-in-the-loop governance, IoT sensing, and blockchain-based accountability.

*3.2. Threat Model*

We assume that an adversary is able to compromise external communication and manipulate nodes that are not authorized. The adversary may attempt:

- **Sensor Spoofing:** Injection of fabricated environmental readings [14].
- **Data Tampering:** Modification of anomaly metadata prior to ledger commitment [15]
- **Alert Injection:** Unauthorized attempt to broadcast public alerts [16].
- **Denial-of-Service (DoS):** Disruption of communication among system components [17].

We suppose that standard cryptographic primitives (digital signatures and hash functions) are secure, as well as the fact that the permissioned blockchain can withstand Byzantine behavior by less than a third of the validator nodes [18].

Human validator private keys are also supposed to be stored in secure hardware (e.g. HSMs) or credential vaults. Revocation procedures are put in place in case of a key compromise being suspected with an on-chain update of any validator credential. Single-point credential compromise can also be reduced by multi-signature authorization schemes [19].

*3.2.1. Human Oracle Risk*

The framework assumes authorized validators act within institutional policy but may still produce incorrect judgments. However, incorrect human approval (false positive or false negative) may still occur. While blockchain ensures non-repudiation, it does not guarantee correctness of human judgment. To mitigate oracle risk, multi-signature approval policies, cross-agency validation, or delayed broadcast with secondary review may be incorporated. These mechanisms reduce reliance on single-operator judgment while preserving accountability [20].

*3.3. Operational Assumptions*

We make assumptions of a limited communication delay $\Delta$, a high enough density of UAVs coverage to achieve periodic coverage of an area, and semi-trusted human validators acting according to institutional protocols. The assumptions are realistic deployment constraints, and they would facilitate sound coordination and governance implementations. With the above assumptions, state-transition rules to blockchain smart contract are cryptographically enforced to prevent unauthorized dissemination of alerts.

## 4. Proposed Governance-Constrained Agentic Framework

We present an agentic governance-responsive monitoring architecture of wildfires based on a combination of limited decision-theoretic modeling, hierarchical multi-agent control, and blockchain-enforced human control in a single control loop.

*4.1. Constrained Decision Formulation*

Wildfire monitoring is developed using a constrained POMDP model but since the exact solutions of the CPOMDP model are computationally infeasible we use a simple heuristic policy based on the constrained model, defined by the tuple ($\mathcal{S}, \mathcal{A}, \mathcal{T}, \mathcal{R}, \omega, \mathcal{O}$). At time *t*, the system state $\mathcal{S}_t \in \mathcal{S}$ is defined as:

$$\mathcal{S}_t = \{\mathcal{H}_t, \mathcal{W}_t, \mathcal{D}_t, \mathcal{R}_t\}$$

where $\mathcal{H}_t$ denotes spatial heat distribution, $\mathcal{W}_t$ meteorological conditions, $\mathcal{D}_t$ UAV configuration states, and $\mathcal{R}_t$ dynamic wildfire risk estimates.

Action space $\mathcal{A}$ includes UAV redeployment coordinates, multi-stage verification triggers, and alert request generation. The observation space $\omega$ consists of multi-modal sensor inputs $\mathcal{O}_t \in \omega$ governed by the observation model $\mathcal{O}$ ($\mathcal{O}_t \mid \mathcal{S}_t, \mathcal{A}_{t-1}$). State transitions follow the dynamic probability model $\mathcal{T}(\mathcal{S}_{t+1} \mid \mathcal{S}_t, \mathcal{A}_t)$. Due to the fact that ignition cannot be directly observed, it is said that the system maintains a belief state:

$$b_t = P(\mathcal{S}_t \mid \mathcal{O}_{1:t}, \mathcal{A}_{1:t-1})$$

The objective is to find a policy $\pi$ that minimizes the expected cost function $\mathcal{R}$:

$$min_\pi \mathbb{E}_\pi[\alpha L_d + \beta F_p + \gamma C_r]$$

where $L_d$ is detection latency, $F_p$ false public alerts, and $C_r$ resource cost.

Optimization is faced with operational constraints (energy and bandwidth) and the following strict governance conditions:

$$Alert_t = 1 \Leftrightarrow (Confidence_t > \tau) \wedge (HumanApproval_t = 1)$$

This inserts human authorization into the action space that is feasible.

*4.2. Hierarchical Multi-Agent Coordination*

We follow a hierarchical architecture in order to solve the constrained decision problem efficiently. Local UAV agents are used to perform both navigation and sensing operations in accordance to a given assigned patrol region, and a centralized meta-controller is used to maintain global belief state b_t and do risk-weighted allocation of tasks:

$$\max_{A_t} \sum_{i=1}^{N} \mathbb{E}[\Delta R(p_i^t)]$$

subject to energy and communication limits. This design allows dynamic coverage of high risk areas whilst keeping the operational cost limited.

*4.3. Multi-Stage Probabilistic Verification*

To reduce false alerts, anomaly validation proceeds in two stages. First, cross-modal fusion produces an anomaly confidence score:

$$\boldsymbol{Confidence_t = f(H_t, W_t) > \tau_1}.$$

Second, verification UAVs collect independent evidence to update the confidence:

$$\boldsymbol{Confidence_t^{final} = g(Confidence_t, V_t)}.$$

Human review is triggered if $\boldsymbol{Confidence_t^{final} > \tau_2, where \tau_2 > \tau_1}$. Thresholds adapt to environmental volatility, balancing sensitivity and specificity. The governance-trigger escalation logic is illustrated in Fig. 2.

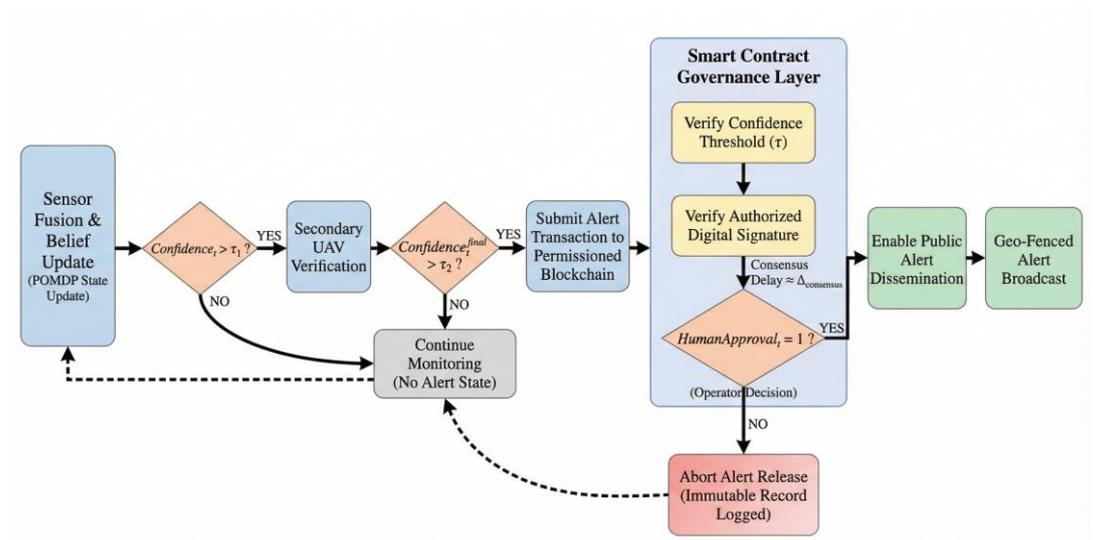

**Figure 2.** Governance-constrained alert authorization flow. Human approval is enforced as a smart-contract state-transition condition prior to public alert dissemination. Negative decisions return to the monitoring loop while preserving immutable audit records.

*4.4. Blockchain-Enforced Human Oversight*

The implementation of governance is through a permissioned blockchain layer that is run by authorized agencies. The anomaly events are hashed and serialized then are committed in form of a transaction. A smart contract does the transition of the state:

$$Alert_t = 1 \Leftrightarrow (Confidence_t > \tau) \wedge (HumanApproval_t = 1).$$

Human approval is presented in the form of the cryptographically signed authorization of an authorized validator. The anomaly threshold and signature authenticity are verified by the smart contract and then alert dissemination is possible. This changes human supervision to procedural protection to a system invariant.

*4.5. Secure Alert Dissemination*

Approved alerts generate a geo-fenced payload:

$$AlertPayload_t = \{GeoBoundary_t, Severity_t, AdvisoryText_t\}.$$

Alerts are shared over the redundancy of communication channels to make it resilient in case of partial infrastructure failure. The level of severity is proposed AI-based and confirmed by humans before a broadcast.

With embedded constrained POMDP optimization, adaptive multi-agent coordination and blockchain-based human validation in one loop, the framework will result into responsive wildfire detection and structural governance safeguards.

## 5. Theoretical Guarantees and Resilience Analysis

Under standard assumptions—specifically, inter-component delay bounded by $\Delta$, secure cryptographic primitives, and fewer than $f < \frac{k}{3}$ Byzantine validators—the architecture provides formal operational guarantees.

**Guarantee 1 (Enforced Authorization)** ensures public alert dissemination strictly satisfies $Alert_t = 1 \Leftrightarrow (Confidence_t > \tau) \wedge (HumanApproval_t = 1)$. Since it is a smart-contract transition, there is no alert broadcastable unless it corresponds to human confirmation which is cryptographically verified.

**Guarantee 2 (Alert Integrity)** implies that the injection or alteration of alerts is computationally infeasible by an unauthorized actor unless adversary passes the Byzantine threshold.

**Proposition 1 (Operational Latency Bound)** establishes that for N UAVs of velocity v monitoring area A, the expected detection latency is bounded by $\mathbb{E}[L_d] \leq \frac{A}{vN} + \Delta$, where accounts for communication overhead. Lastly, the events of all anomalies and human approvals are signed digitally, hashed and stored on chain permanently. This ensures non-repudiation, replay attacks are prevented by unique identifiers and nonces and operational continuity is ensured.

## 6. Experimental Evaluation

In this section, the performance of detection, the strength of the governance, and scalability of the suggested governance-constrained agentic structure are evaluated. The entire system, such as multi-agent coordination, blockchain validation, and HITL authorization, was presented as executable modules in a simulated controlled environment.

*6.1. Experimental Setup*

The wildfire environment was simulated to be a 100 × 100 grid with vegetation, humidity, wind, and temperature. Fire propagation followed a stochastic cellular automaton:

$$P_{spread} = \sigma(\alpha_1 Wind + \alpha_2 Fuel - \alpha_3 Humidity).$$

Each scenario ran for T=3000 time steps (10 seconds per step).

**Coordination Engine:** The belief states were represented in the form of probabilistic heat maps which were updated through Bayesian filtering. Risk-weighted allocation solved:

$$\max_{z \in Z} R(z) \cdot Coverage(z),$$

using greedy approximation $(O(N|\mathcal{Z}|))$. This heuristic does not solve the optimal CPOMDP solution, but instead is an approximation of the constrained objective by risk-weighted allocation but explicitly applying governance constraints at the state-transition level.

**Sensing Modalities:**

- Thermal UAV sensors (0.85 detection probability),
- Ground IoT sensors,
- Satellite feed with bounded latency.

Synthetic anomalies were injected on the signals and non-fire sources of heat.

**Blockchain and HITL:**

A permissioned blockchain with k = 7 validators (tolerating f = 2 Byzantine nodes) enforced:

$$Alert_t = 1 \Leftrightarrow (Confidence_t > \tau) \wedge (HumanApproval_t = 1).$$

Average validation delay: 1.2 time steps. This latency is based on a permissioned network that is locally deployed and has consistent connectivity. Consensus latency can grow significantly in geographically dispersed networks of disasters, or in degraded networks. The architecture would be correct even in the face of increased delay, but alert release time would increase with network conditions.

Human review delay: mean 3 time steps. Each of the experiments was repeated on 20 random seeds.

*6.2. Baselines*

- Proposed (Agentic + Blockchain)
- Adaptive AI (No Governance)
- Static Monitoring

These isolate coordination, governance, and oversight effects.

*6.3. Metrics*

- Detection latency $(L_d)$
- False alert rate $(F_p)$
- Human override frequency
- Blockchain confirmation delay

*6.4. Results*

*6.4.1. Human Oracle Risk*

Fig. 3 depicts that the latency reduces inversely with the UAV density, which is in line with the theoretical limit.

Enforcement of governance produces an insignificant latency compared to adaptive AI without governance (<5% increase across all densities).

Static patrol exhibits limited scalability. Error bars represent standard deviation across 20 simulation seeds.

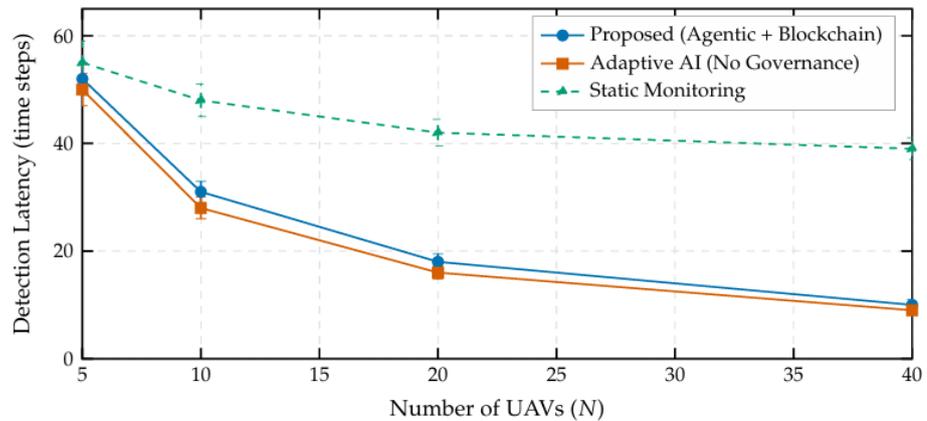

**Figure 3.** Detection latency comparison across baselines with standard deviation error bars. Adaptive coordination dominates static monitoring, while governance enforcement introduces negligible latency overhead.

*6.4.2. False Alert Reduction*

The adaptive base is more aggressive in dealing with false alerts because the products of aggressive anomaly thresholding without multi-stage checks are high, whereas the product of interaction is lower in dealings with the static patrol, which is less responsive and less inclined towards over-triggering. As Fig. 4 shows, false alerts are significantly reduced in the case of synthetic anomaly injection. The adaptive AI baseline (without governance) can achieve 22% false alert rate, whereas the proposed framework has 6% ($p < 0.01$). The statistical significance was determined by paired two-sided t-test on 20 independent simulation seeds. Traceability is enhanced by blockchain logging but not by false dissemination, which necessitates obligatory HITL.

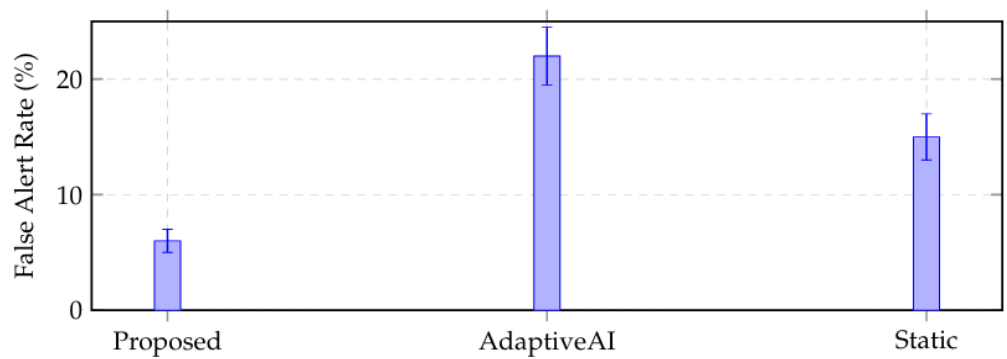

**Figure 4.** False public alert rate under synthetic anomaly injection. The proposed governance-constrained framework consistently reduces erroneous alert dissemination relative to static and unguided adaptive baselines.

*6.4.3. Performance–Governance Tradeoff*

Fig. 5 illustrates the latency–false alert frontier. The proposed framework achieves Pareto dominance: low latency that is similar to adaptive AI at a significantly lower false alert.

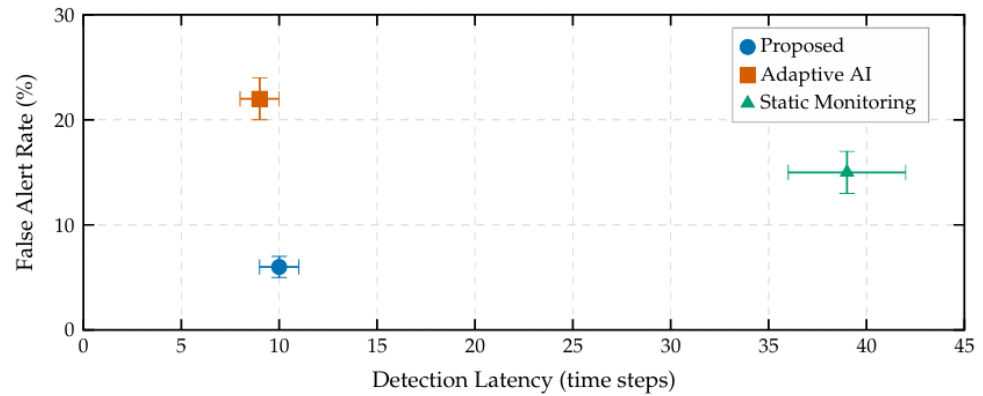

**Figure 5.** Latency-false alert tradeoff across monitoring architectures. The proposed framework achieves lower false alert rates while maintaining competitive detection latency.

*6.4.4. Governance Overhead*

Even in the case of high anomaly frequency, blockchain confirmation delay is limited (1–3 time steps).

To quantify the latency–governance trade-off, we decomposed end-to-end alert delay into:

1. sensing and verification latency,
2. coordination latency,
3. blockchain consensus delay, and
4. human validation delay.

Under nominal load, blockchain confirmation contributes approximately 12% of total alert latency, while human review contributes 28%. During simulated high-alert bursts (5× anomaly frequency), blockchain latency increased by 35% but remained below sensing-induced delay. This indicates that governance enforcement, while non-negligible, does not dominate the critical alert path.

Governance-related delay was not more than 8% of total detection latency in all UAV densities, which confirms that sensing and patrol dynamics are the most appropriate aspects of system responsiveness in the conditions studied.

*6.4.5. Ablation Study*

The elimination of adaptive coordination adds 30-45% of latency. The elimination of HITL raises false alerts over 3 times. When blockchain enforcement is removed, the performance of detection is maintained, verifiable audit trails are removed, and it is possible to simulate transitions to alert states that cannot be authorized under adversarial conditions.

*6.4.6. Scalability*

Latency is proportional to the area under surveillance and inverse to the number of UAVs. For fixed validator count (k=7), consensus overhead remains effectively constant. Assuming that the number of validators scales with the size of deployment, the communication complexity of BFT is also of standard size (typically, it is proportional to $O(k^2)$), and can raise coordination latency.

Overall, it has been shown that governance constraints are implementable in adaptive multi-agent control without compromising detection efficiency, and that they enhance integrity and reliability of public alerts dramatically.

# 7. Conclusion

The paper introduced a governance-restricted agentic system of safety-critical wildfire monitoring with human supervision that is enforced by blockchains. The model of wildfire detection was the constrained POMDP that combined the latency, minimization of false alarms, and resource minimization with the explicit authorization limitations. Hierarchical multi-agent coordination mechanism facilitated the adaptive redeployment of the UAVs, whereas smart contracts were used to impose a state-transition invariant of human approval before the public disclosure.

With a set of assumptions of bounded communication delay and Byzantine fault, the framework ensures authorization is enforced, alert integrity, non-repudiation, and a limited operational latency. Simulation findings indicate that there is a small overhead and a huge decrease in false public alerts in governance enforcement compared to autonomous baselines.

The results indicate that adaptive autonomy and enforceable accountability may be structurally combined in a single control loop, which will turn human oversight into a procedural protection instead of an operational system constraint. Large scale inference, consensus overhead, infrastructure disruption and perception-layer robustness are issues which should be considered for practical deployment. Adversarial robust sensing, formalization of governance logic, and validation of the scale of a field will be addressed in future work. In general, the incorporation of governance as a control invariant offers a principled basis of accountable safety-critical AI systems.